\input harvmac

\Title{\vbox{\baselineskip12pt\hbox{UICHEP-TH/98-6}}}
{\centerline{Resonant Bound State Production at $e^-e^-$ Colliders}}

\centerline{David~Bowser-Chao\footnote{$^1$}{davechao@hepster.phy.uic.edu},
Tom~D.~Imbo\footnote{$^2$}{imbo@uic.edu},
B.~Alex~King\footnote{$^3$}{aking@uic.edu} and
Eric~C.~Martell\footnote{$^4$}{ecm@uic.edu}}

\bigskip\centerline{Department of Physics}
\centerline{University of Illinois at Chicago}
\centerline{845 W.  Taylor St.}
\centerline{Chicago, IL \ 60607-7049}
\vskip 0.75in

Observation of a sequence of resonances at an $e^-e^-$ collider would suggest bound states of strongly coupled constituents carrying lepton number. Obvious candidates for these exotic constituents are leptoquarks and leptogluons. We show that under reasonable assumptions, the existence of one leptogluon flavor of appropriate mass can give rise to sizeable ``leptoglueball''  production rates and observable resonance peaks. In contrast, one needs two leptoquark flavors in order to produce the analogous ``leptoquarkonium'' states. Moreover, cross-generational leptoquark couplings are necessary to give observable event rates in many cases, and leptoquarkonium mass splittings are too small to resolve with realistic beam energy resolutions. 

\Date{}

Numerous TeV-scale colliders have been proposed as possible complements
to the planned Large Hadron Collider (LHC). For example,
$e^+e^-$, $\mu^+\mu^-$, $e^-\gamma$, $\gamma\gamma$ and $e^-e^-$ beams are all being
seriously considered. Because of their lepton number $L = 2$ initial
state, $e^-e^-$ colliders are generally thought to have the least
physics potential. However such a facility could, for example, study
strong $W^-W^-$ scattering --- an aspect of electroweak
symmetry breaking not readily accessible at the LHC or other machines
\ref\ewsb{H. Murayama, Int.~J.~Mod.~Phys.~A {\bf 13} (1998) 2227\semi T. Han,
Int.~J.~Mod.~Phys.~A {\bf 13} (1998) 2337.}. Moreover, if doubly-charged particles with $L=2$
exist and have reasonable couplings to electrons, they would be produced
copiously at an $e^-e^-$ collider through an $s$-channel resonance~\ref\bilep{P. Frampton, Int.~J.~Mod.~Phys.~A {\bf 13} (1998) 2345.}\ref\bilepp{M. Raidal,
Phys. Rev. D {\bf 57} (1998) 2013\semi F. Cuypers and M. Raidal, Nucl. Phys. B {\bf 501}
(1997) 3.}. If such
{\it bileptons} are fundamental (or have a compositeness scale which is
well above the collider energy range), then there will be a single resonance peak. If, however, these bileptons are composite at these energy scales, then there should be a sequence of peaks representing excitations of the underlying constituents, allowing a spectroscopic study similar to well-known treatments of quarkonium systems.

Because they carry lepton number and have strong interactions, {\it colored 
leptons} provide a possible mechanism for the formation of such
bound states. For example, {\it leptogluons} ($LG$s) can occur in quark/lepton composite models containing colored
preons~\ref\lgs{H. Fritzsch and G. Mandelbaum, Phys. Lett. {\bf 102B} (1981) 319\semi H.Harari and N. Seiberg,
Phys. Lett. {\bf 98B} (1981) 269\semi E. J. Squires, Phys. Lett. {\bf 94B} (1980) 54.}\ref\lgss{T. G. Rizzo, Phys. Rev. D {\bf 34} (1986) 133;
Phys. Rev. D {\bf 33} (1986) 1852\semi Y.~Nir, Phys.~Lett. {\bf 164B} (1986) 395\semi K.~H.~Streng, Z.~Phys. C {\bf 33} (1986) 247\semi U.~Baur and K.~H.~Streng, Z.~Phys. C {\bf 30} (1986) 325; Phys.~Lett. {\bf 162B} (1985) 387\semi H.~Harari, Phys.~Lett. {\bf 156B} (1985) 250\semi S.~F.~King and S.~R.~Sharpe, Nucl. Phys. B {\bf 253} (1985) 1; Phys.~Lett. {\bf 143B} (1984) 494.}\ref\hewriz{J. L. Hewitt and
T. G. Rizzo, Phys. Rev. D {\bf 56} (1997) 5709.}. $LG$s would be color octets, like Standard Model gluons, but would
also carry lepton number, electric charge, and have a spin of 1/2 or 3/2. They would have all the
couplings of ordinary leptons and gluons, as well as additional
trilinear couplings to lepton-gluon pairs. Current experiments tell us that, if they exist, spin-1/2 $LG$s which couple to $e^-g$ must have masses $m_{\scriptscriptstyle LG}{\ \lower-1.2pt\vbox{\hbox{\rlap{$>$}\lower5pt\vbox{\hbox{$\sim$}}}}\
}325$~GeV \hewriz. We will show that a single flavor of such $LG$ will give rise, under reasonable assumptions, to sizeable production rates of $LG$-$LG$ bound states (leptoglueballs) at an $e^-e^-$ collider. Moreover, even given the poor beam energy resolution expected at such a machine, the bound state mass splittings are big enough (and widths small enough) so that some resonance peaks can likely be reconstructed. 

The relevant (non-renormalizable) interaction term in the lagrangian for spin-1/2 $LG$s has the form \hewriz
\eqn\lag{{g_s \over \Lambda} \left[\lambda_L{\bar e_R}\sigma^{\mu\nu}G^a_L + \lambda_R {\bar\ell_L} \sigma^{\mu\nu}G^a_R \right] F^a_{\mu\nu} +  h.c.\,\,\,\, .} 
Here ${\bar\ell_L}$ is the left-handed first generation $SU(2)_L$
lepton doublet, $G_R^a$ the right-handed $LG$ doublet, $e_R$ and $G^a_L$
the corresponding singlets, $F^a_{\mu\nu}$ the $SU(3)_c$ field strength
tensor, and $g_s=\sqrt{4\pi\alpha_s(\mu)}$ the QCD running coupling. Defining the
dimensionless couplings $\lambda_L$ and $\lambda_R$ to be of order 1,
$\Lambda$ represents the ``compositeness scale'' of the $LG$. In order to avoid possible problems with magnetic moment constraints, we assume a purely chiral interaction. Specifically, we take $\lambda_R = 0$ and $\lambda_L=1$ so that the leptogluon only couples to the right-handed electron $e_R$. (The choice $\lambda_R=1$, $\lambda_L=0$ gives equivalent results for the charge $-1$ component of the right-handed $LG$ doublet.) If the leptogluon and electron are simply different color states of the same underlying preon configuration, then $m_{\scriptscriptstyle LG}$ will be ${\cal O}(\Lambda\alpha_s(\Lambda ))$ \lgs .  

We now discuss leptoglueball production, restricting our attention to $S$-wave \hbox{($\ell = 0$)} states. (The production of $P$-wave states is suppressed relative to these by a factor of order~$\alpha_s^2$.) Since these leptoglueballs are formed from identical fermions, their
overall wavefunctions must be antisymmetric under the simultaneous
interchange of color, spin and spatial variables. Hence, for color
singlet $S$-states, the two $LG$s must be in the antisymmetric spin-0 state. (This is analogous to the situation for bound states of two
gluinos \ref\yee{W.-Y.~Keung and A.~Khare, Phys. Rev. D {\bf 29} (1984) 2657.}.) There are two primary decay channels for these leptoglueballs, namely the annihilation of the two $LG$s (which occurs via $t$-channel gluon exchange)
\eqn\gamggee{\Gamma_{B\to e^-_R
e^-_R}=288\pi\,\alpha_s(m_{\scriptscriptstyle LG})^2\,\left|\psi(0)\right|^2{M_B^2\over\Lambda^4},}
and the spectator decay of one of the $LG$s
\eqn\gamegX{\Gamma_{B\to e^-_R g + X}=2\Gamma_{LG\to e^-_R
g}=2\,\alpha_s(m_{\scriptscriptstyle LG})\,{m_{\scriptscriptstyle LG}^3 \over \Lambda^2}.}
Here $M_B$ is the mass of the nonrelativistic bound state $B$ and $\left|\psi(0)\right|^2$ the square of its wavefunction at the origin. Since these $LG$s must be very heavy, it is reasonably accurate to describe leptoglueballs using the perturbative Coulombic potential \hbox{$V(r) = -{C_2 \alpha_s(\mu) \over r}$}. (We will ignore hyperfine effects.) For the $nS$ state this leads to \hbox{$M_B\simeq 2m_{\scriptscriptstyle LG}-\left( {C_2\alpha_s(\mu_n)\over 2n}\right)^2\!\! m_{\scriptscriptstyle LG}$} and $\left|\psi (0)\right|^2\simeq {1\over \pi}\left({C_2 \alpha_s(\mu_n)\over 2 n} \,m_{\scriptscriptstyle LG}\right)^3$. The ``Casimir factor'' $C_2$ is 3 for color octets (compared to 4/3 for color triplets), and \hbox{$\mu_n={C_2  \alpha_s(m_{\scriptscriptstyle LG}) \over 2 n}\,m_{\scriptscriptstyle LG}$} is a typical momentum scale for the $nS$ state. For all channels this gives \hbox{$\Gamma_{B\to e^-_R e^-_R} \ll 
\Gamma_{B\to e^-_R g + X}$}, and the total width is well-approximated as \hbox{$\Gamma (B) \simeq \Gamma_{B\to e^-_R g + X}.$}

Using the above results, we can now compute the total production cross section, $\sigma_{\hbox{tot}}$, for $e^-_{\scriptscriptstyle R}e^-_{\scriptscriptstyle R}\to B$:
\eqn\sigsmear{\sigma_{\hbox{tot}} = {32 \pi\over M_B^2} {\Gamma_{B\to e^-_R e^-_R} \over \Gamma (B)} \cdot f(x),}
where $f(x) = 2x\,e^{x^2}\int_x^{\infty}e^{-t^2}dt$ and $x={1\over\sqrt{8}}{\Gamma (B)\over\sigma_E}$.
The center-of-mass energy has been assumed to have a gaussian distribution peaked at the resonance mass $M_B$ with a spread of $\sigma_{\scriptscriptstyle E}$. The factor $f$ takes into account the effects of this beam energy uncertainty: $f(x)\to 1$ as $\sigma_{\scriptscriptstyle E}\to 0$ and $f(x)\simeq\sqrt{\pi}\,x$ for $\sigma_{\scriptscriptstyle E}\gg \Gamma (B)$. 
For $nS$ states in the above Coulombic approximation, \sigsmear\ becomes simply
\eqn\gamcoulomb{\sigma_{\hbox{tot}} = 15552\pi \cdot
{\alpha_s(m_{\scriptscriptstyle LG})^3
\alpha_s(\mu_n)^3\over m_{\scriptscriptstyle
LG}^2n^3}\cdot f(x),}  where we have set $\Lambda = {m_{\scriptscriptstyle LG}
\over  \alpha_s(m_{\scriptscriptstyle LG})}$. The expected beam energy resolution for a \hbox{TeV-scale} $e^-e^-$ collider is \hbox{$\sigma_{\scriptscriptstyle E}\simeq 0.01\sqrt{s}{\ \lower-1.2pt\vbox{\hbox{\rlap{$>$}\lower5pt\vbox{\hbox{$\sim$}}}}\ } 10\,\Gamma (B)$}, giving an $f(x)$ on the order of 0.05.

We have used \gamcoulomb\ to calculate the expected number of events for
the production of 1$S$ leptoglueballs on resonance (see Table~1). The
event rates for the 2$S$ (respectively, 3$S$) states will be smaller by about a factor of
6 (respectively, 16). Note that these calculations did not require any unnatural assumptions about the strength of the coupling to produce sizeable rates. These 
rates should be several orders of magnitude larger than the QED
background $e^- e^- \to e^-e^- + (\gamma^*\to$~2~jets).  In addition, since
a leptoglueball (at rest) decaying into $e^- e^- +$~2 jets will look
very different from this QED process, the application of
appropriate cuts should reduce this background even further. In the Coulombic approximation, the mass splitting between the 1$S$ and 2$S$ states ($\Delta_{2S-1S}$) ranges from
\hbox{$\sim$ 10} to 20~GeV as $m_{\scriptscriptstyle LG}$ varies from
400 to 1000 GeV. This is about 14 times bigger than the natural width $\Gamma (B)$ of the states for all values of $m_{\scriptscriptstyle LG}$.  ($\Delta_{3S-2S}$ is about $3\,\Gamma (B)$.) Moreover, the distance between the 1$S$ resonance peak and the continuum $LG$ pair production threshold at $\sqrt{s}=2 m_{\scriptscriptstyle LG}$ is $\sim 20\,\Gamma (B)$. Thus if we had unlimited beam resolution, at least the 1$S$ and 2$S$ resonance peaks should be easily visible. However, the expected beam energy spread $\sigma_{\scriptscriptstyle E}$ is about the same as $\Delta_{2S-1S}$. Even so, it may still be possible to reconstruct the $1S$ state, as well as some evidence for the $2S$ state, from a detailed analysis of the shape of the observed cross section. Obtaining a $\sigma_{\scriptscriptstyle E}$ which is a factor of 5 to 10 smaller than expected would not only go a long way in helping us to see these successive peaks, but would also increase the production cross sections by increasing $f(x)$.\foot{One can also consider the production of ``second generation'' leptogluons at a $\mu^-\mu^-$ collider, which is expected to have a much better beam energy resolution of $\sim 2\times 10^{-5}\sqrt{s}$~\ref\demarteau{M. Demarteau and T. Han, hep-ph/9801407.}.} 
\vskip 18pt
\vbox{\offinterlineskip
$$\vbox{
\halign{\strut\vrule\quad\hfil#\hfil\quad&\vrule\quad\hfil#\hfil\quad\vrule\cr
\noalign{\hrule}
\vrule height11pt depth3.5pt width0pt $m_{\scriptscriptstyle LG}$&$e^-_{\scriptscriptstyle R} e^-_{\scriptscriptstyle R} \rightarrow B$\cr
\noalign{\hrule}
400&$1.472 \times 10^5$\cr
\noalign{\hrule}
600&$4.267 \times 10^4$\cr
\noalign{\hrule}
800&$1.796 \times 10^4$\cr
\noalign{\hrule}
1000&$9.243 \times 10^3$\cr
\noalign{\hrule}
}}$$
\vskip 6pt
\centerline{\vbox{\hsize=10cm\noindent
{\bf Table 1:} Estimates$\strut$ of the number of $1S$ leptoglueball events on resonance at a polarized $e^-e^-$ collider. $\strut$The numbers are obtained from \gamcoulomb\ using the three-loop$\strut$ formula for $\alpha_s(\mu)$ given in~\ref\LQpaper{D.~Bowser-Chao, T.~D.~Imbo, B.~Alex~King and E.~C.~Martell, Phys. Lett. B {\bf 432} (1998) 167.}.  $\strut$Masses are in GeV and the integrated luminosity is taken to be $\strut$10~fb$^{-1}$.$\strut$}}}
\vskip 12pt

It is interesting to compare the above leptoglueballs with {\it fundamental} scalar bileptons $S$ of the same mass and quantum numbers.  If we denote the (dimensionless) trilinear coupling of $S$ to two right-handed electrons by $g$, then we would like to know the ``effective'' $g$ for leptoglueballs. That is, for what value of $g$ does $\Gamma_{B\to e^-_R e^-_R}$ from \gamggee\ equal \hbox{$\Gamma_{S\to e^-_{\scriptscriptstyle R} e^-_{\scriptscriptstyle R}}={g^2M_S\over 8\pi}$ \bilepp?} For example, setting $M_S=M_B=800$~GeV these two widths are equal (for 1$S$ leptoglueballs) if $g\simeq {1\over 100}$, where again we have used the Coulombic approximation and $\Lambda = {m_{\scriptscriptstyle LG} \over  \alpha_s(m_{\scriptscriptstyle LG})}$. Thus, the effective coupling of leptoglueballs to electron pairs is small but non-negligible.

So far, the focus of this letter has been the resonant production of composite bileptons at $e^- e^-$ colliders with spin-1/2 leptogluons as the constituents. Of course, we can also consider the production of these hypothetical $LG$s at other future colliders. For example, the best scenario for the {\it discovery} of $LG$s would most likely be single production at an $e^-p$ machine. There are also lepton-number-zero analogs of our leptoglueballs, namely \hbox{$LG$-$\overline {LG}$} bound states, which can in principle be produced on resonance at an $e^+e^-$ collider. Ignoring extremely small electroweak effects, spin-zero $LG$-$\overline {LG}$ states with $\ell =0$ will have the same spectrum as our $LG$-$LG$ states above, although with somewhat larger widths due to additional annihilation decay channels.  However, the cross section for resonant production of these scalar bound states at an $e^+e^-$ machine is proportional to $(\lambda_L\lambda_R)^2$, requiring both $\lambda_L$ and $\lambda_R$ in \lag\ to be non-zero. The measured electron magnetic moment then constrains $\lambda_L\lambda_R$ to be very small, resulting in a huge suppression.  Thus, the spin-0 $LG$-$LG$ states that we have considered in the context of $e^-e^-$ collisions have no observable $LG$-$\overline {LG}$ analog (on resonance) at an $e^+e^-$ collider. It {\it will} be possible to produce spin-1 $LG$-$\overline {LG}$ bound states with $\ell =0$ on resonance in $e^+e^-$ collisions (with reasonably high rates and low backgrounds). Interestingly, these states have no analog in $LG$-$LG$ systems due to the exclusion principle. So we see that $e^+ e^-$ and $e^- e^-$ colliders will allow us to study complementary aspects of the spectroscopy of bound states of leptogluons. 

Finally, we will briefly consider another candidate for the constituents of composite bileptons, namely {\it leptoquarks} ($LQ$s) --- color triplets (or antitriplets) carrying lepton number and having a spin of either 0 or 1. These objects have been the subject of much discussion
over the last few years (see, for example, the papers cited in
\ref\lqhistory{P. H. Frampton and M. Harada, hep-ph/9711448.}).
However the resonant production of a bound
state of two $LQ$s (leptoquarkonia) at an $e^- e^-$ collider is somewhat unnatural, requiring two distinct leptoquark flavors:  one which couples to $e^- q$ and one which couples to $e^- \bar q$ for some quark flavor $q$.\foot{Production of bound states containing leptoquarks at other future colliders has been discussed in \LQpaper. For a discussion
of continuum leptoquark production at $e^-e^-$ colliders, see
\ref\cuyp{F.~Cuypers, P.~Frampton and R.~R\"uckl, Phys. Lett. B {\bf
390} (1997) 221.}.} In addition, many such leptoquarkonium production rates are ``quark mass suppressed'' due to a necessary chirality flip of the exchanged $t$-channel quark. Examples of this are the production of an $S$-wave bound state of two scalar $LQ$s or of one scalar and one vector $LQ$. In such cases, a cross-generational coupling to the top quark is necessary to produce non-negligible event rates. The cross section for the production of $S$-wave bound states of two {\it vector} $LQ$s from incoming electrons with opposite helicity does {\it not} have the above quark mass suppression. ($P$-wave bound states can also be produced without a chirality flip regardless of the $LQ$ spin.) But in any case, even if the above leptoquarkonium states do exist and have reasonable production rates, the weaker QCD interactions of color triplets versus color octets will lead to bound state mass splittings about 4 times smaller than those of leptoglueballs of similar mass. Since these splittings are much smaller than the expected beam energy resolution, it is unlikely that any resonance peaks could be reconstructed. All we are likely to see is a broad enhancement of the continuum $LQ$ pair production threshold, with no obvious sign of a resonance peak.


In conclusion, we have seen that the existence of spin-1/2 leptogluons of appropriate mass which couple to $e^- g$ could give rise to
striking bound state resonances (leptoglueballs) at a TeV-scale $e^- e^-$ collider.  These bileptonic resonances would arise more naturally and be much easier to observe than those produced by leptoquarks of similar mass. The production of a pair of identical 1 TeV $LG$s will lead, under reasonable assumptions, to 1$S$ and 2$S$ scalar leptoglueball resonance peaks well above background, separated by $\sim$~20~GeV, each having a width of about 1.4~GeV. The 2$S$ peak, which is $\sim$ 1/6 the height of the 1$S$ peak, remains about 9~GeV below the continuum $LG$ pair production threshold. One hopes that a sufficiently small beam resolution can be achieved in order to allow us to resolve these peaks. This analysis is sensitive only to the assumption that $\Lambda
={m_{\scriptscriptstyle LG}\over \alpha_s(m_{\scriptscriptstyle LG})}$.
Although this seems natural from a composite model point of view, 
\gamggee --\sigsmear\ tell us that any significant deviation from this equality would greatly affect both $\Gamma (B)$ and $\sigma_{\hbox{tot}}$ (while not changing the bound state mass splittings).

\vskip12pt

\noindent 

\centerline{\bf Acknowledgements}

\vskip 6pt

\noindent
It is a pleasure to acknowledge Lee Brekke, Adam Falk, Howard Georgi, Tao Han, \hbox{Wai-Yee~Keung}, \hbox{Young-Kee Kim}, Joe Lykken, Uday Sukhatme, Scott Willenbrock and Dieter Zeppenfeld for helpful discussions. A very special thank you goes to Phil Jackson for years of guidance. This research was supported in part by the U.S. Department of Energy under Grant Number DE-FG02-91ER40676.

\listrefs

\raggedbottom

\bye